\documentclass[
prd
,showpacs,amssymb,superscriptaddress,aps
]{revtex4}
\usepackage{graphicx}
\usepackage{color}
\input{epsf}

\usepackage{amsmath,amssymb}
\usepackage{bm}
\usepackage{times}
\usepackage{ulem}

\newcommand{\dalm}{\kern1pt\vbox{\hrule height 0.9pt\hbox{\vrule width 0.9pt
\hskip 2.5pt\vbox{\vskip 5.5pt}\hskip 3pt\vrule width 0.3pt}\hrule height 0.3pt}
\kern1pt}

\newcommand{\lsim}{\, \, \raisebox{-0.8ex}{$\stackrel{\textstyle <}{\sim}$ }}

\begin{document}



\title{Gravitational wave asteroseismology for low-mass neutron stars}

\author{Hajime Sotani}
\email{sotani@yukawa.kyoto-u.ac.jp}
\affiliation{Astrophysical Big Bang Laboratory, RIKEN, Saitama 351-0198, Japan}
\affiliation{Interdisciplinary Theoretical \& Mathematical Science Program (iTHEMS), RIKEN, Saitama 351-0198, Japan}


\date{\today}

\begin{abstract}
The fundamental ($f$-) mode gravitational waves from cold low-mass neutron stars are systematically studied with various equations of state (EOSs) characterized by the nuclear saturation parameters, especially focusing on the phenomena of the avoided crossing with the first pressure ($p_1$-) mode. We find that the $f$-mode frequency and the average density for the neutron star at the avoided crossing can be expressed as a function of the parameter, $\eta$, which is the specific combination of the nuclear saturation parameters. Owing to these relations, we can derive the empirical formula expressing the $f$-mode frequency for a low-mass neutron star, whose central density is larger than that for the neutron star at the avoided crossing, as a function of $\eta$ and the square root of the stellar average density, $x$. On the other hand, we also derive the empirical formula expressing the $f$-mode frequency for a neutron star, whose central density is less than that for the neutron star at the avoided crossing, as a function of $x$ independently of the adopted EOS. Furthermore, adopting the empirical formula of $x$ as a function of $\eta$ and $u_c$, which is the ratio of the stellar central density to the saturation density, we can also rewrite our empirical formula for the $f$-mode frequency to a function of $\eta$ and $u_c$. So, by observing the $f$-mode gravitational wave from a low-mass neutron star, whose mass or gravitational redshift is known, one could evaluate the values of $\eta$ and $u_c$, which enables us to severely constrain the EOS for neutron star matter. 
\end{abstract}

%
\maketitle


\section{Introduction}
\label{sec:I}

Neutron stars provided via the supernova explosions are a suitable laboratory for seeing the physics under the extreme conditions. In fact, the density inside the star significantly exceeds the standard nuclear density, and the gravitational and magnetic fields inside/around the star become much stronger than those observed in the solar system. Via the constraints on the neutron star properties by observing the neutron star itself and/or the phenomena associated with the neutron stars, one would extract the information about such extreme conditions. For example, the discoveries of the $2M_\odot$ neutron stars are enable us to exclude some of soft equations of state (EOSs) \cite{D10,A13,C20}. In addition, the properties of the millisecond pulsar PSR J0030+0451 could observationally be estimated by the Neutron star Interior Composition Explorer (NICER) mission \cite{Riley19,Miller19}. This is because the light curves from a rotating neutron star with a hot spot mainly depend on the compactness of neutron star, which is the ratio of the mass to the radius, as a result of the light bending due to a relativistic effect (e.g., \cite{PFC83,LL95,PG03,PO14,SM18,Sotani20}).

As another approach to observationally extract the neutron star properties, asteroseismology is also powerful technique, which is similar to seismology on the Earth and helioseismology on the Sun. That is, since the oscillation frequencies strongly depend on the interior properties of objects, one can extract the {\it invisible} information via observation of such frequencies as an inverse problem. In practice, by identifying the quasi-periodic oscillations observed in the giant flares with the crustal torsional oscillations in the neutron star, the crustal properties can be constrained \cite{GNHL2011,SNIO2012,SIO2016}. In a similar way, via the observation of gravitational waves from the compact objects, it is proposed that one would get the information about the stellar mass, radius, and EOS for a high density region (e.g., \cite{AK1996,AK1998,STM2001,SH2003,SYMT2011,PA2012,DGKK2013}), which is sometimes referred to as gravitational wave asteroseismology. Moreover, with respect to the gravitational waves from the supernovae, it is also discussed how the gravitational wave signals appearing in the numerical simulations correspond to the  specific gravitational wave modes in the protoneutron stars (e.g., \cite{FMP2003,FKAO2015,ST2016,SKTK2017,MRBV2018,TCPOF19,SS2019}). Thanks to the success of the direct detection of gravitational waves and the electromagnetic counterparts from the binary neutron star merger \cite{GW6,EM}, now gravitational waves really become a new tool to see the astronomical information, with which one may be able to practice asteroseismology on the compact objects some day.

The gravitational waves from the (cold) neutron stars have various modes \cite{KS1999}. According to the input physics, the corresponding modes can be excited. Thus, if one would observe a specific mode in gravitational waves, one could inversely see the counterpart in physics. From the observational point of view, the fundamental ($f$-) and the first (and possibly the second) pressure ($p_i$-) modes may be important, because those frequencies are relatively low among various modes (although their frequencies are still more than kilohertz). Since the $f$-mode is a kind of acoustic oscillations, its frequency is considered to be characterized by the stellar average density, which is $M/R^3$ with the stellar mass $M$ and radius $R$. In fact, it is shown that the $f$-mode frequency is written as a linear function of the square root of the stellar average density, where the dependence on the adopted EOS is weak \cite{AK1996,AK1998}. Nevertheless, if the small dependence on the adopted EOS in the linear relation between the $f$-mode frequency and the square root of the average density could be described with a specific parameter characterizing the EOS, one would extract the EOS information via the observation of the $f$-mode gravitational waves. For this purpose, in this study we systematically examine the $f$-mode frequency in low-mass neutron stars constructed with various EOSs, focusing on the nuclear saturation parameters as the parameter characterizing the EOS. Since the density inside the low-mass neutron stars is relatively not so high, the stellar properties are considered to be directly associated with the nuclear saturation parameters. In fact, it is shown that some of the neutron star properties are written well as a function of the stellar central density and the suitable combination of the saturation parameters, $\eta$ \cite{SIOO14,SSB16}. In a similar way, in this study we will try to derive the empirical formula for the $f$-mode frequency from the low-mass neutron stars as a function of the stellar central density and $\eta$.

This paper is organized as follows. In Sec. \ref{sec:NS}, we describe the EOSs and low-mass neutron star models considered in this study. In Sec. \ref{sec:GW}, we show the eigenfrequencies of gravitational waves from the low-mass neutron stars, where we especially  focus on the properties at the avoided crossing between the $f$- and $p_1$-modes. Then, we derive the empirical formula for the $f$-mode frequency. Finally, we make a conclusion in Sec. \ref{sec:Conclusion}. Unless otherwise mentioned, we adopt geometric units in the following, $c=G=1$, where $c$ denotes the speed of light, and the metric signature is $(-,+,+,+)$.

\section{EOS parameters and Low-mass Neutron star models}
\label{sec:NS}

In this study, we simply consider the spherically symmetric stellar models, which are constructed by integrating the Tolman-Oppenheimer-Volkoff equations together with an appropriate EOS. Various EOSs have been proposed up to now, but the EOS for neutron star matter is not fixed yet. This mainly comes from two reasons, i.e., the lack of observational constraints on the neutron stars and the difficulty for obtaining the nuclear information in high density region from terrestrial experiments. On the other hand, owing to the nature of nuclear saturability, the EOS in lower density region, such as around the saturation density, is gradually constrained via terrestrial experiments. In fact, any EOSs are characterized by the nuclear saturation parameters. That is, the bulk energy, $w$, of nuclear matter with zero temperature for any EOSs can be expanded in the vicinity of the saturation point as a function of the baryon number density, $n_{\rm b}$, and neutron excess, $\alpha$, such as
\begin{equation}
  w = w_0 +  \frac{K_0}{18n_0^2}(n_{\rm b}-n_0)^2 + \left[S_0 + \frac{L}{3n_0}(n_{\rm b}-n_0)\right]\alpha^2,
     \label{eq:w}
\end{equation}
where $w_0$ and $K_0$ denote the bulk energy and incompressibility for the symmetric nuclear matter at the saturation density, $n_0$, while $S_0$ and $L$ are the parameters associated with the nuclear symmetry energy, $S(n_{\rm b})$, via $S_0=S(n_0)$ and $L = 3n_0(dS/dn_{\rm b})$ at $n_{\rm b}=n_0$. Among these five saturation parameters, $n_0$, $w_0$, and $S_0$ are well constrained from the experimental data for masses and charge radii of stable nuclei, but the constraint on the remaining two parameters, $K_0$ and $L$, is relatively more difficult, because these are associated with the derivative of $n_{\rm b}$ at $n_{\rm b}=n_0$, i.e., one has to prepare the experimental data in the wide range of $n_{\rm b}$ around $n_{\rm b}=n_0$ in order to constrain $K_0$ and $L$. Thus, in this study, we focus on $K_0$ and $L$ to see the dependence of the gravitational wave frequency on the EOS parameters. For this purpose, we especially adopt the phenomenological EOS constructed by Oyamatsu and Iida \cite{OI03,OI07} (hereafter we referee to this EOS as OI-EOSs) together with the Shen EOS based on the relativistic mean field theory \cite{Shen}. We remark that one can choose the values of $K_0$ and $L$ in OI-EOSs, where the other parameters are fixed in such a way as to recover the experimental data for stable nuclei. The concrete EOS parameters considered in this study are listed in Table \ref{tab:EOS}, where $\eta$ is a parameter defined by Eq. (\ref{eq:eta}) and we also show the transition density, $n_{\rm c}$, from the crust (composed of non-uniform matter) to core (composed of uniform matter).

\begin {table}
\caption{EOS parameters adopted in this study, where the transition density, $n_{\rm c}$, between the crust and core is also shown. }
\label{tab:EOS}
\begin {center}
\begin{tabular}{cccccc}
\hline\hline
EOS & $K_0$ (MeV) & $L$ (MeV) & $\eta$ (MeV) & $n_{\rm c}$ (fm$^{-3}$)   \\
\hline
OI-EOSs
 &   180 & 31.0   & 55.8  & 0.09068  &  \\
 &   180 & 52.2   & 78.9  & 0.07899  &  \\
 &   230 & 42.6   & 74.7  & 0.08637  &  \\
 &   230 & 73.4   & 107   & 0.07345  &  \\
 &   280 & 54.9   & 94.5  & 0.08331  &  \\
 &   280 & 97.5   & 139   & 0.06887  &  \\  
\hline
Shen
 & 281 & 111  & 151  &  0.058  & \\
\hline
\end{tabular}
\end {center}
\end{table}

\begin{figure}[tbp]
\begin{center}
\includegraphics[scale=0.5]{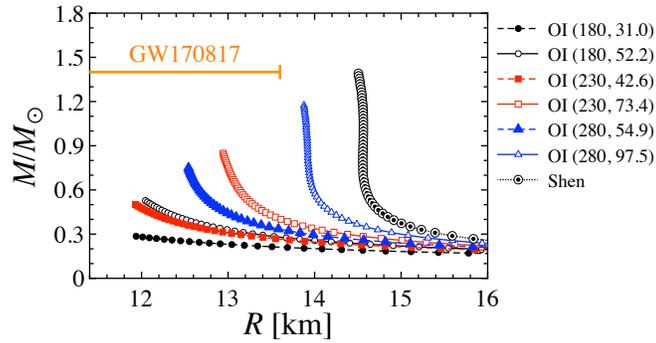}  
\end{center}
\caption{
Mass-radius relation with various EOSs for $\rho_c\le 2\rho_0$. For reference, we also show the radius of a $1.4M_\odot$ neutron star constrained from GW170817.
}
\label{fig:MR}
\end{figure}

Since the neutron star structure in high density region depends strongly on the adopted EOSs, where one may have to take into account the many-body interactions and the contribution from non-nucleonic components such as hyperons and quarks, in this study we consider only the low-mass neutron stars, whose central density is less than $2\rho_0$ with the saturation density $\rho_0=2.68 \times 10^{14}$ g cm$^{-3}$. In fact, in this density region, one can avoid the contribution from non-nucleonic components \cite{LP04}. In addition, according to the quantum Monte Carlo calculations, the uncertainty from three-neutron interactions in pure neutron matter is not so relevant in the density region of $\rho\lsim 2\rho_0$ \cite{GCR12}. In Fig. \ref{fig:MR}, we show the relation between the stellar mass, $M$, and radius, $R$, for the low-mass neutron stars constructed with the EOSs listed in Table \ref{tab:EOS}, where for reference we also show the radius constraint obtained from the event GW170817, i.e., the maximum radius of a $1.4M_\odot$ neutron star should be less than 13.6 km \cite{Annala18}. With this constraint, some of the EOSs considered in this study may be ruled out. Even so, in order to examine the dependence of the gravitational wave frequencies in the wide parameter range, we consider the EOSs listed in Table \ref{tab:EOS}.

It has been shown that such low-mass neutron stars can be characterized well by a new parameter, $\eta$, defined as \cite{SIOO14}
\begin{equation} 
  \eta = (K_0 L^2)^{1/3}.   \label{eq:eta}
\end{equation}
In practice, the mass, $M$, and gravitational redshift, $z$, (which consequently reduces to the radius) of low-mass neutron stars are expressed as a function of $\eta$ and $u_c\equiv \rho_c/\rho_0$, i.e., $M=M(\eta,u_c)$ and $z=z(\eta,u_c)$, where $\rho_c$ is the central density of neutron stars. In a similar way, the moment of inertia, quadrupole moment, quadrupole ellipticity, tidal and rotational Love number, and apsidal constant of slowly rotating low-mass  neutron stars, can be expressed as a function of $\eta$ and $u_c$ \cite{SSB16}. Furthermore, by assuming that the EOS characterized by $\eta$ is adopted in the density region of $\rho\le 2\rho_0$ and the causal limit EOS, i.e., the sound velocity is the same as the speed of light, is adopted in the density region of $\rho\ge 2\rho_0$, the possible maximum mass of neutron stars is given by 
\begin{equation}
   \frac{M_{\rm max}}{M_\odot} = 2.856 + 0.1511 \eta_{100}, \label{eq:Mmax}
\end{equation}
where $\eta_{100}\equiv \eta/100\ {\rm MeV}$ \cite{Sotani17}.

\begin{figure}[tbp]
\begin{center}
\includegraphics[scale=0.5]{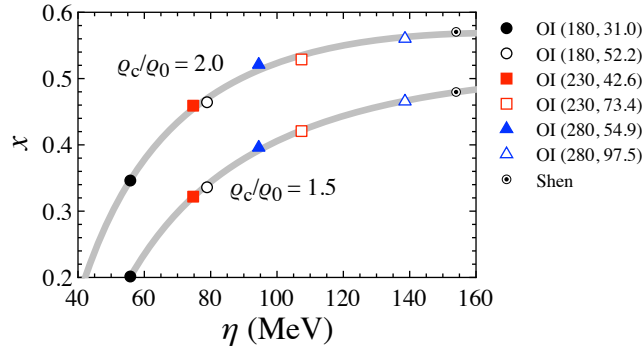}  
\end{center}
\caption{
For $\rho_c/\rho_0=1.5$ and $2.0$, the square root of the normalized stellar average density, $x\equiv (M/1.4M_\odot)^{1/2}(R/10\ {\rm km})^{-3/2}$, constructed by various EOSs is shown as a function of $\eta$. The thick-solid line denotes the fitting line with Eq. (\ref{eq:x_eta}).
}
\label{fig:etax}
\end{figure}

\begin{figure}[tbp]
\begin{center}
\includegraphics[scale=0.5]{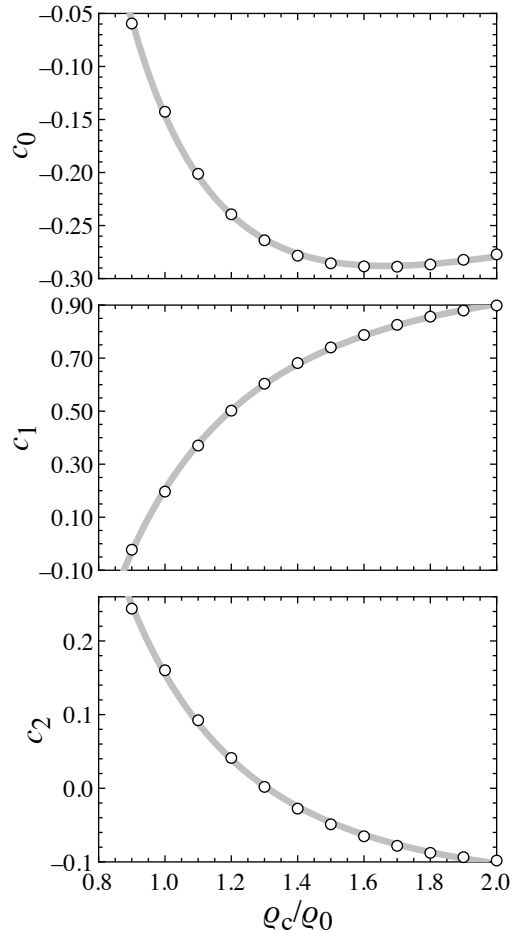}  
\end{center}
\caption{
The coefficients in Eq. (\ref{eq:x_eta}) are shown as a function of $u_c\equiv \rho_c/\rho_0$. The thick-solid line in each panel is fitting line given by Eqs. (\ref{eq:c0}) - (\ref{eq:c2}).
}
\label{fig:c012}
\end{figure}

As in Refs. \cite{SIOO14,SSB16}, we additionally find that the square root of the normalized average density, $x\equiv (M/1.4M_\odot)^{1/2}(R/10\ {\rm km})^{-3/2}$, for the low-mass neutron stars with the fixed central density can be described as a function of $\eta$. In Fig. \ref{fig:etax} we show the value of $x$ for $u_c=1.5$ and $2.0$ constructed with various EOSs are shown as a function of $\eta$, where the thick-solid lines are fitting lines given by
\begin{equation}
  x = c_0\eta_{100}^{-1} + c_1 + c_2\eta_{100}. \label{eq:x_eta}
\end{equation}
In this formula, $c_0$, $c_1$, and $c_2$ are coefficients depending on $u_c$. To see the dependence of these coefficients on $u_c$, we examine the fitting lines given by Eq. (\ref{eq:x_eta}) as varying the value of $u_c$. Then, the resultant coefficients are shown in Fig. \ref{fig:c012} as a function of $u_c$, where the top, middle, and bottom panels correspond to $c_0$, $c_1$, and $c_2$, respectively. In this figure, the thick-solid lines are fitting lines given by
\begin{gather}
  c_0(u_c) =  0.8885u_c^{-2} -1.0670u_c^{-1} +  0.03225, \label{eq:c0} \\
  c_1(u_c) =  -1.1810u_c^{-2} +0.3742u_c^{-1} + 1.0120, \label{eq:c1} \\
  c_2(u_c) =  0.5375u_c^{-2} -0.2919u_c^{-1} - 0.09055. \label{eq:c2}
\end{gather}
At last, we derive the empirical formula (Eq. (\ref{eq:x_eta}) with Eqs. (\ref{eq:c0}) - (\ref{eq:c2})) describing $x$ for low-mass neutron stars as a function of $\eta$ and $u_c$ as well as $M$ and $z$, i.e., $x=x(\eta,u_c)$.

\section{GW asteroseismology}
\label{sec:GW}

\begin{figure}[tbp]
\begin{center}
\includegraphics[scale=0.5]{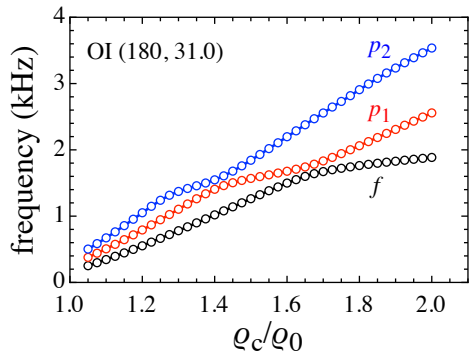}  
\end{center}
\caption{
For the low-mass neutron star constructed with OI-EOSs ($K_0=180$ and $L=31.0$ MeV), the frequencies of the $f$-, $p_1$-, and $p_2$-modes are shown as a function of the central density normalized by the saturation density.
}
\label{fig:f-y350k180}
\end{figure}

On the low-mass neutron stars discussed in the previous section, we discuss the gravitational wave frequencies obtained by linear analysis. In this study, we simply adopt the relativistic Cowling approximation, i.e., the metric perturbations are neglected during the fluid oscillations. The perturbation equations are derived by linearizing the energy-momentum conservation law. Then, with appropriate boundary conditions at the center and surface of neutron star, the problem to solve becomes an eigenvalue problem. The concrete perturbation equations and boundary conditions are the same as in Ref. \cite{SYMT2011}. Since the neutron star models considered in this study are constructed with zero temperature EOS without any density discontinuities, the excited oscillations are only the $f$- and $p_i$-modes. That is, in our model, the $f$-mode frequency is the lowest frequency theoretically expected. In addition, we focus on only the $\ell=2$ modes in this study, because it is considered that the $\ell=2$ modes would be the most energetic signal among the modes for $\ell \ge 2$. As an example, in Fig. \ref{fig:f-y350k180} we show the $f$-, $p_1$-, and $p_2$-mode frequencies as a function of $u_c$ for the stellar model constructed with OI-EOSs with $K_0=180$ and $L=31.0$ MeV. From this figure, one can clearly observe the phenomena of the avoided crossing between the eigenmodes, i.e., the $f$- and $p_1$-modes around $u_c\simeq 1.65$ and the $p_1$- and $p_2$-modes around $u_c\simeq 1.4$ \cite{note}. We henceforth focus on the $f$-mode frequency, because this mode must be observationally the most important gravitational wave signals from the compact object. We remark that the typical $f$-mode frequency from low-mass neutron stars becomes around kHz, which must be difficult for detecting with the current gravitational wave detectors. We hope the next generation detector(s), such as Einstein telescope, would probe the $f$-mode frequencies from low-mass neutron stars, although it may be still challenging.

\begin{figure}[tbp]
\begin{center}
\includegraphics[scale=0.5]{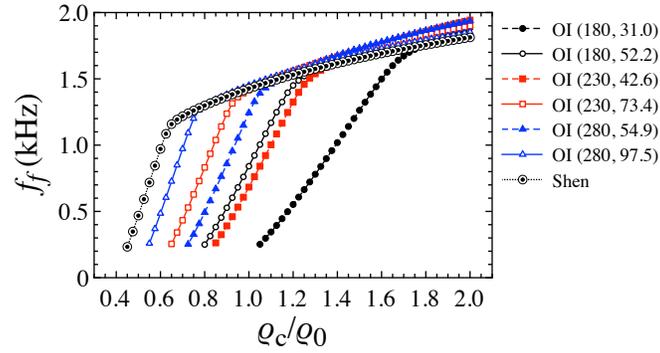}  
\end{center}
\caption{
For low-mass neutron stars constructed with various EOSs, the $f$-mode frequency is shown as a function of the central density normalized by the saturation density.
}
\label{fig:ff-ratio}
\end{figure}

\begin{figure}[tbp]
\begin{center}
\includegraphics[scale=0.5]{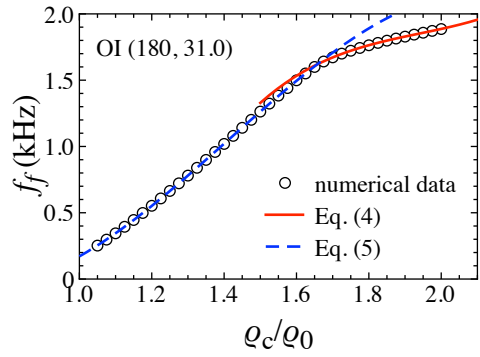}  
\end{center}
\caption{
With OI-EOSs ($K_0= 180$ and $L=31.0$ MeV), the $f$-mode frequencies before and after the avoided crossing are respectively fitted with Eqs. (\ref{eq:fit_fa}) and (\ref{eq:fit_fb}).
}
\label{fig:ff-fit}
\end{figure}

\begin{figure*}[tbp]
\begin{center}
\includegraphics[scale=0.5]{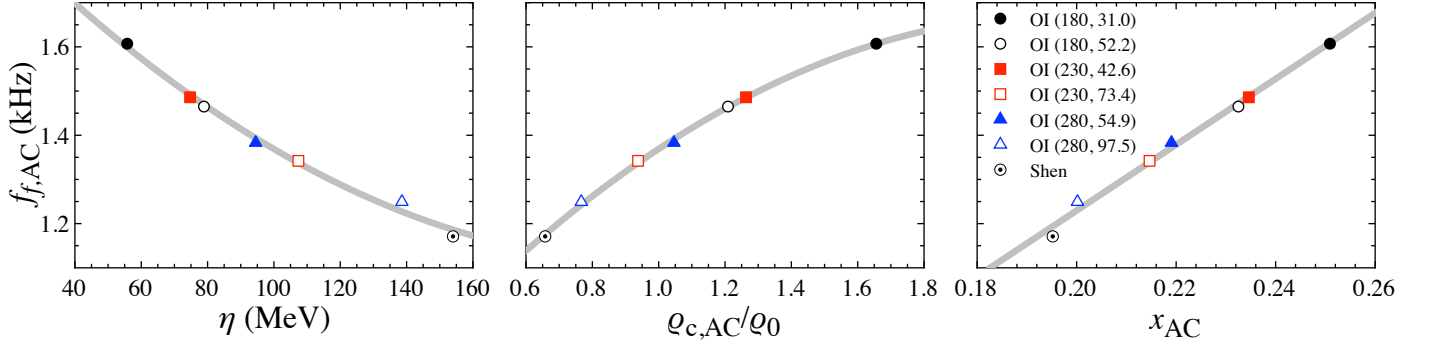}
\end{center}
\caption{
The $f$-mode frequencies at the avoided crossing, $f_{f, {\rm AC}}$, for various EOSs are shown as a function of $\eta$ (left), $\rho_{c,{\rm AC}}/\rho_0$ (middle), and $x_{\rm AC}$ (right), where $x_{\rm AC}$ denotes the value of $x\equiv (M/1.4M_\odot)^{1/2}(R/10\ {\rm km})^{-3/2}$ for the neutron star model at the avoided crossing between the $f$- and $p_1$-modes. In each panel, the thick-solid line denotes the fitting formulae given by Eqs. (\ref{eq:fAC-eta}), (\ref{eq:fAC-ratio}), and (\ref{eq:fAC-x}) respectively.
}
\label{fig:ffAC}
\end{figure*}

\begin{figure}[tbp]
\begin{center}
\includegraphics[scale=0.5]{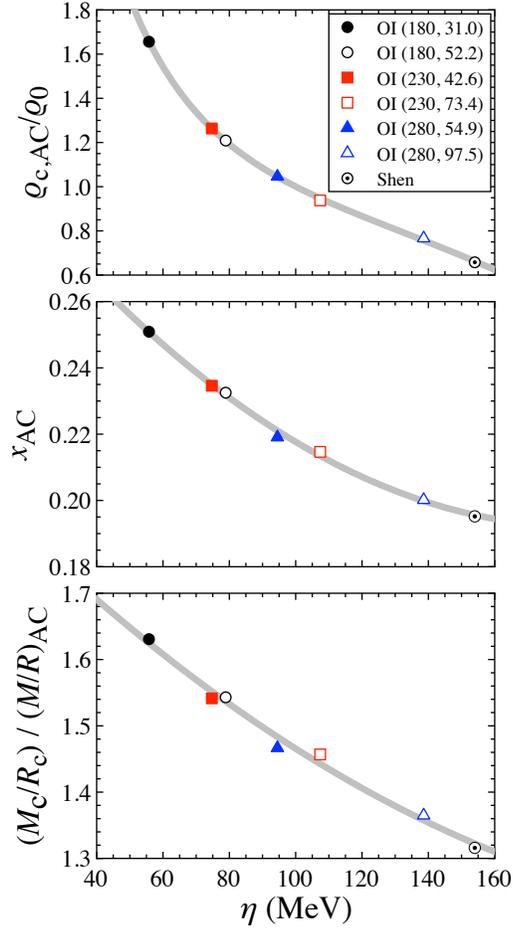}  
\end{center}
\caption{
The values of $\rho_{c,{\rm AC}}/\rho_0(\equiv u_{c,{\rm AC}})$ (top), $x_{\rm AC}$ (middle), and $(M_{\rm c}/R_{\rm c})/(M/R)_{\rm AC}$ (bottom) for various EOSs are shown as a function of $\eta$. The thick-solid line denotes the fitting formulae given by Eqs. (\ref{eq:ratio-eta}), (\ref{eq:x-eta}), and (\ref{eq:comp-eta}),  respectively. Note that $(M_{\rm c}/R_{\rm c})/(M/R)_{\rm AC}$ is the ratio of the compactness for the core region, $M_{\rm c}/R_{\rm c}$, to stellar compactness, $M/R$, for the neutron star model at the avoided crossing between the $f$- and $p_1$-modes, where $M_{\rm c}$ and $R_{\rm c}$ denote the mass and radius of the core region of neutron star. 
}
\label{fig:eta-ratioAC}
\end{figure}

In Fig. \ref{fig:ff-ratio} we show the $f$-mode frequencies for the low-mass neutron stars constructed with various EOSs. From this figure, one can see that the central density of the neutron star, with which the avoided crossing between the $f$- and $p_1$-modes happens, depends strongly on the adopted EOS. In order to discuss the properties on the avoided crossing between the $f$- and $p_1$-modes, as in Fig. \ref{fig:ff-fit}, the $f$-mode frequencies before and after the avoided crossing are respectively fitted as a cubic function of $u_c$ as
\begin{gather}
  f_f ({\rm kHz}) = a_1 + a_2 u_{c} + a_3 u_{c}^2 + a_4 u_{c}^3, \label{eq:fit_fa} \\
  f_f ({\rm kHz}) = b_1 + b_2 u_{c} + b_3 u_{c}^2 + b_4 u_{c}^3, \label{eq:fit_fb} 
\end{gather}
where $a_i$ and $b_i$ with $i=1-4$ are fitting coefficients depending on the adopted EOS, and then the $f$-mode frequency and the neutron star model at the avoided crossing are determined as the intersection between Eqs. (\ref{eq:fit_fa}) and (\ref{eq:fit_fb}). With respect to the resultant $f$-mode frequency, $f_{f,{\rm AC}}$, we find that $f_{f, {\rm AC}}$ can be expressed well as a function of $\eta$, $u_{c,\rm {AC}}\equiv \rho_{c,{\rm AC}}/\rho_0$, or $x_{\rm AC}$, where $\rho_{c,{\rm AC}}$ and $x_{\rm AC}$ denote the central density and $x$ for the neutron star model at the avoided crossing between the $f$- and $p_1$-modes, respectively. In Fig. \ref{fig:ffAC}, we plot $f_{f,{\rm AC}}$ as a function of $\eta$ (left), $u_{c,\rm {AC}}$ (middle), and $x_{\rm AC}$ (right) together with the fitting formulae given by 
\begin{gather}
  f_{f,{\rm AC}}\ ({\rm kHz}) = 1.9926 -0.8095 \eta_{100} + 0.1856 \eta_{100}^2,  \label{eq:fAC-eta}  \\
  f_{f,{\rm AC}}\ ({\rm kHz}) = 0.6718 + 0.8995 u_{c,{\rm AC}} - 0.2022 u_{c,{\rm AC}}^2, \label{eq:fAC-ratio} \\
  f_{f,{\rm AC}}\ ({\rm kHz}) = -0.2608 + 7.4522 x_{\rm AC}. \label{eq:fAC-x}
\end{gather}
So, if one would observationally see the $f$-mode frequency at the avoided crossing, one could extract the value of $\eta$, $u_{c,{\rm AC}}$, and $x_{\rm AC}$. Additionally, we find that not only $u_{c,{\rm AC}}$ and $x_{\rm AC}$ but also the ratio of the compactness of core region, $M_{\rm c}/R_{\rm c}$, to the stellar compactness, $M/R$, for the neutron star model at the avoided crossing can be expressed well as a function of $\eta$, where $M_{\rm c}$ and $R_{\rm c}$ denote the mass and radius of core region. We remark that $R_{\rm c}$ is the radial position, where $n_{\rm b}$ is equal to $n_{\rm c}$ shown in Table \ref{tab:EOS}, and $M_{\rm c}$ is the mass inside $R_{\rm c}$. In Fig. \ref{fig:eta-ratioAC}, we plot $u_{c,{\rm AC}}$ (top), $x_{\rm AC}$ (middle), and $(M_{\rm c}/R_{\rm c})/(M/R)$ (bottom) for the neutron star model at the avoided crossing as a function of $\eta$, where we also shown the fitting formulae given by 
\begin{gather}
  u_{c,{\rm AC}} = 1.2012 \eta_{100}^{-1} - 1.1616 +1.4840 \eta_{100} - 0.5233 \eta_{100}^2,  
  \label{eq:ratio-eta} \\
  x_{\rm AC} = 0.3120 - 0.1291 \eta_{100} +0.034754 \eta_{100}^2,  \label{eq:x-eta} \\
  (M_{\rm c}/R_{\rm c})/(M/R)_{\rm AC} = 1.8794 - 0.5104 \eta_{100} + 0.096551 \eta_{100}^2.  \label{eq:comp-eta}
\end{gather}
With Eqs. (\ref{eq:fAC-eta}) and (\ref{eq:comp-eta}), one may extract the information for the boundary between the core and crust region inside the star via the detection of the $f$-mode frequency for the neutron star at the avoided crossing, where $\eta$ is an intervening variable \cite{crust}.

\begin{figure}[tbp]
\begin{center}
\includegraphics[scale=0.5]{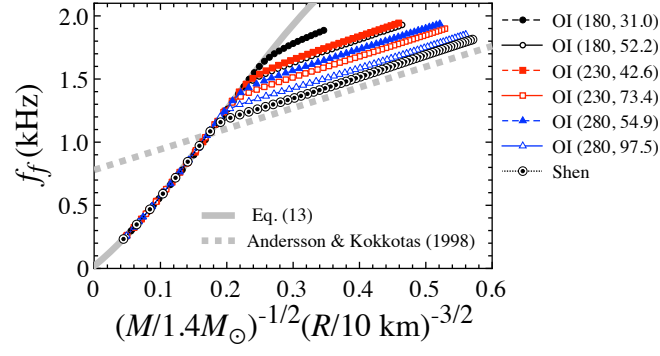}  
\end{center}
\caption{
Same as in Fig. \ref{fig:ff-ratio}, but as a function of square root of the normalized average density, $(M/1.4M_\odot)^{1/2}(R/10\ {\rm km})^{-3/2}$. The thick-solid line denotes the fitting formula for the $f$-mode frequency before the avoided crossing given by Eq. (\ref{eq:ff-before}), while the thick-dotted line denotes the fitting formula proposed by Andersson and Kokkotas given by Eq. (\ref{eq:ff-AK}) \cite{AK1998}.
}
\label{fig:ff-ave}
\end{figure}

Next, we consider the $f$-mode frequency as a function of $x$. Since the $f$-mode (and $p_i$-modes) is a kind of acoustic oscillations, its frequency is characterized by the sound speed, which is associated with the stellar average density.  In fact, it has been shown that the $f$-mode frequency is expressed well as a linear function of $x$ \cite{AK1998}, such as
\begin{equation}
  f_f ({\rm kHz}) = 0.78 + 1.635 x. \label{eq:ff-AK}
\end{equation}
In a similar way, in Fig. \ref{fig:ff-ave} we show the $f$-mode frequency for the low-mass neutron stars constructed with various EOSs. From this figure, one can observe that the $f$-mode frequencies for the neutron star whose central density is less than $\rho_{c,{\rm AC}}$ are independent from the adopted EOSs, which is expressed well as a function of $x$ as
\begin{equation}
  f_f ({\rm kHz}) = 0.018687 + 4.2621x + 15.1014x^2 - 26.8770x^3. \label{eq:ff-before}
\end{equation}
In Fig. \ref{fig:ff-ave} we also show the frequencies calculated with Eq. (\ref{eq:ff-before}) with the thick-solid line. Meanwhile, for a  neutron star model, whose central density is larger than $\rho_{c,{\rm AC}}$, the $f$-mode frequency seems to be more or less expressed as Eq. (\ref{eq:ff-AK}). In fact, the slope of the linear behavior between the $f$-mode frequency and $x$ is similar to that in Eq.  (\ref{eq:ff-AK}) independently of the adopted EOSs. Nevertheless, the dependence of the $f$-mode frequency on the EOSs may not be negligible, where the $f$-mode frequencies deviate within $\sim 0.5$ kHz (which corresponds to $\sim 30\%$ difference).

\begin{figure}[tbp]
\begin{center}
\includegraphics[scale=0.5]{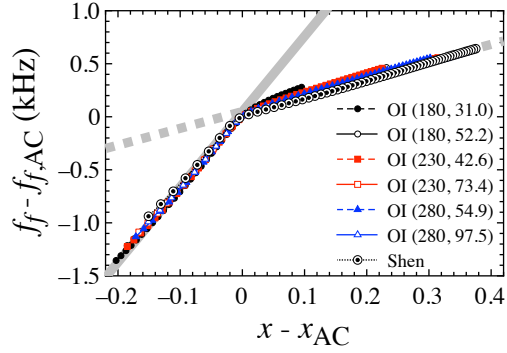}  
\end{center}
\caption{
$f_f-f_{f,{\rm AC}}$ is shown as a function of $x-x_{\rm AC}$ for various EOSs. The thick-solid and thick-dotted lines denote the fitting formulae given by Eqs. (\ref{eq:fit1}) and (\ref{eq:fit2}), respectively.
}
\label{fig:ff-x}
\end{figure}

To derive the empirical formula for the $f$-mode frequency with more accurate, the $f$-mode frequencies at the avoided crossing are collected at the same point, as in Fig. \ref{fig:ff-x}. In this figure, the thick-solid and thick-dotted lines are fitting formula for $f_f-f_{f,{\rm AC}}$ in the range of $x\le x_{\rm AC}$ and $x\ge x_{\rm AC}$, respectively, which are given by 
\begin{gather}
  f_f -f_{f,{\rm AC}}\  ({\rm kHz}) = 7.1036(x-x_{\rm AC}) + 0.049128, \label{eq:fit1} \\
  f_f -f_{f,{\rm AC}}\ ({\rm kHz}) = 1.5834(x-x_{\rm AC}) + 0.049128. \label{eq:fit2}
\end{gather}
Furthermore, since $f_{f,{\rm AC}}$ and $x_{\rm AC}$ are expressed as a function of $\eta_{100}$ as Eqs. (\ref{eq:fAC-eta}) and (\ref{eq:x-eta}), Eq. (\ref{eq:fit2}) can be rewritten as a function of $x$ and $\eta_{100}$ as
\begin{gather}
  f_f\ ({\rm kHz}) = 1.5834x + 0.1306\eta_{100}^2 - 0.6051\eta_{100} + 1.5477. \label{eq:fit3}
\end{gather}
At last, we can obtain the empirical formula for the $f$-mode frequency. That is, the $f$-mode frequencies before and after the avoided crossing are expected by Eqs. (\ref{eq:ff-before}) and (\ref{eq:fit3}), respectively. Moreover, as mentioned in the previous section, $x$ is written as a function of $\eta$ and $u_c$ as Eq. (\ref{eq:x_eta}) with Eqs. (\ref{eq:c0}) - (\ref{eq:c2}). Thus, we can rewirte the empirical formula for $f$-mode frequency excited in low-mass neutron stars to a function of $\eta$ and $u_c$, i.e., $f_f=f_f(\eta,u_c)$. So, if one would observe the $f$-mode gravitational wave from a low-mass neutron star, whose mass or gravitational redshift is known, one could evaluate the values of $\eta$ and the stellar central density. This information must give us a severe constraint on the EOS for neutron star matter.

\begin{figure}[tbp]
\begin{center}
\includegraphics[scale=0.5]{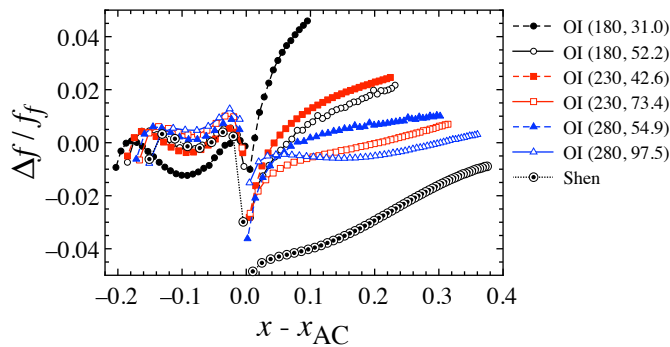}  
\end{center}
\caption{
Relative deviation, $\Delta f/f_f$, of the $f$-mode frequency from the empirical formula given by Eqs. (\ref{eq:ff-before}) and (\ref{eq:fit3}) is shown as a function of $x-x_{\rm AC}$, where $\Delta f/f_f$ is calculated by Eq. (\ref{eq:df}).
}
\label{fig:df-xx}
\end{figure}

Finally, we have to check the accuracy of our empirical formula. For this purpose, we calculate the relative deviation of the $f$-mode frequency from the empirical formula, such as
\begin{equation}
  \frac{\Delta f}{f_f} = 1 - \frac{f_{\rm em}}{f_f}, \label{eq:df}
\end{equation}
where $f_{\rm em}$ denotes the frequency calculated with empirical formula given by Eqs. (\ref{eq:ff-before}) and (\ref{eq:fit3}), while $f_f$ denotes the $f$-mode frequency for each neutron star model. The resultant values are shown in Fig. \ref{fig:df-xx} as a function of $x-x_{\rm AC}$. From this figure, we confirm that our empirical formula can predict the $f$-mode frequency with less than $5\%$ accuracy. Additionally, since the accuracy becomes the worst at the point of the avoided crossing, if one would observe the $f$-mode frequencies from various low-mass neutron stars, one may identify the point of the avoided crossing. 

\begin{figure}[tbp]
\begin{center}
\includegraphics[scale=0.5]{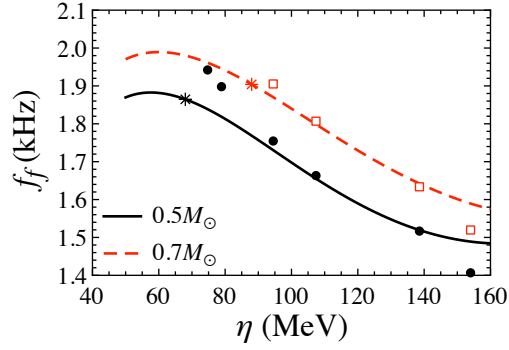}  
\end{center}
\caption{
The $f$-mode frequencies for $0.5M_\odot$ (solid line) and $0.7M_\odot$ (dashed line) neutron star models are shown as a function of $\eta$ with using the methodology proposed in this study, i.e., with using the fitting formulae for the $f$-mode frequency and the mass as a function of $u_c$ and $\eta$. The circles and squares are the $f$-mode frequencies for the neuron star models constructed with the EOS considered in this study. On each line, the asterisk denotes the neutron star model, where $u_c$ becomes 2. Since we derive the fitting formulae by considering the region for $u_c\le 2$, the left-hand part of the asterisk is incredible.
}
\label{fig:fM-eta}
\end{figure}

On the other hand, we also have to check how well the methodology proposed in this study can work. For this purpose, with respect to the neutron star models with  $M=0.5M_\odot$ and $0.7M_\odot$, we compare the $f$-mode frequencies obtained with the fitting formula for the mass and the $f$-mode frequency to those for the neutron star models constructed with the specific EOS considered in this study. That is, once the stellar mass is observed (or fixed), the normalized central density, $u_c$, is determined as a function of $\eta$ via the mass formula (Eq. (2) in Ref. \cite{SIOO14}). Then, via the fitting formula of the $f$-mode derived in this study (Eq. (\ref{eq:fit3}) together with Eqs. (\ref{eq:x_eta}) - (\ref{eq:c2})), one can estimate the $f$-mode frequencies as a function of $\eta$ (or $u_c$). The resultant $f$-mode frequencies are shown as a function of $\eta$ with the solid line for $M=0.5M_\odot$ and the dashed line for $0.7M_\odot$ in Fig. \ref{fig:fM-eta}. On this figure, we also plot the $f$-mode frequencies for the neutron star models constructed with the EOS considered in this study with the circles for $M=0.5M_\odot$ and the squares for $0.7M_\odot$. We remark that since we consider the low-mass neutron stars shown in Fig. \ref{fig:MR}, we do not consider the neutron star models with lower value of $\eta$, whose mass does not approach $0.5M_\odot$ or $0.7M_\odot$. From this figure, we can conclude that our methodology proposed in this study works well, if $\eta$ is in the range of $90\lsim \eta\lsim 140$ MeV. The reason why our methodology does not work well for lower value of $\eta$ may come from the fact that the central density of neutron star with lower value of $\eta$ becomes higher for the neutron star models with fixed mass, while the uncertainty in a high density region becomes significant as discussed in Ref. \cite{SIOO14}, which leads to that the mass and $f$-mode formulae do not work well in such a region. In fact, since we derive the fitting formulae by considering the region for $u_c\le 2$, the line to the left of the asterisk (with lower value of $\eta$) in Fig. \ref{fig:fM-eta} is incredible, where the asterisk denotes the neutron star model with $u_c=2$. In order to complement this inadequacy, one may have to find an additional nuclear parameter expressing the higher density region.

\section{Conclusion}
\label{sec:Conclusion}

 In this paper, we focused on the low-mass neutron stars and examined the frequencies of gravitational waves from such objects. We found that the avoided crossing can generally be observed in the frequencies of gravitational waves, varying the central density of the neutron star, where the central density of the neutron star model at the avoided crossing strongly depends on the adopted EOS. We particularly examined the neutron star models at the avoided crossing between the $f$- and $p_1$-modes for various EOSs and we found that the $f$-mode frequency at the avoided crossing can be expressed as a function of $\eta$, which is a specific combination of the nuclear saturation parameters, the central density, and the square root of the average density of the neutron star models at the avoided crossing. In the same manner, the central density, the square root of the average density, and the ratio of core compactness to the stellar compactness of the neutron star models at the avoided crossing can be expressed as a function of $\eta$. Owing to the relation between the $f$-mode frequency at the avoided crossing and $\eta$ together with the relation between the square root of the average density of the neutron star models at the avoided crossing and $\eta$, we could derive the empirical formula expressing the $f$-mode frequency for a neutron star model, whose central density is larger than that for the neutron star at the avoided crossing, as a function of the square root of the average density and $\eta$ (Eq. (\ref{eq:fit3})), with which the $f$-mode frequency can be estimated with less than $5\%$ accuracy. On the other hand, the $f$-mode frequency from the neutron stars whose central density is less than that for the neutron star at the avoided crossing can be expressed as a function of the stellar average density independently of the adopted EOS (Eq. (\ref{eq:ff-before})). Finally, adopting the empirical relation for the square root of the  average density as a function of $\eta$ and $u_c$, which is the ratio of the stellar central density to the saturation density, we can rewrite the empirical formula for the $f$-mode frequency to a function of $\eta$ and $u_c$. That is, if one would observe the $f$-mode gravitational wave from a low-mass neutron star, whose mass or gravitational wave redshift is known, one could evaluate the values of $\eta$ and $u_c$, because the mass and gravitational wave redshift are also written as a function of $\eta$ and $u_c$. 

\acknowledgments

This work is supported in part by Japan Society for the Promotion of Science (JSPS) KAKENHI Grant Number JP19KK0354 and Ministry of Education, Science and Culture of Japan (MEXT) KAKENHI Grant Numbers JP20H04753.



\begin{thebibliography}{999}

\bibitem{D10} 
   P. Demorest, T. Pennucci, S. Ransom, M. Roberts, and J. Hessels, Nature {\bf 467}, 1081 (2010).

\bibitem{A13} 
   J. Antoniadis {\it et al.}, Science {\bf 340}, 6131 (2013).

\bibitem{C20}    
   H. T. Cromartie {\it et al.}, Nature Astronomy {\bf 4}, 72 (2020).

\bibitem{PFC83} 
   K. R. Pechenick, C. Ftaclas, and J. M. Cohen, Astrophys. J. {\bf 274}, 846 (1983).

\bibitem{LL95} 
   D. A. Leahy and L. Li, Mon. Not. R. Astron. Soc. {\bf 277}, 1177 (1995).


\bibitem{PG03} 
   J. Poutanen and M. Gierlinski, Mon. Not. R. Astron. Soc. {\bf 343}, 1301 (2003).

\bibitem{PO14} 
   D. Psaltis and F. \"{O}zel, Astrophys. J. {\bf 792}, 87 (2014). 

\bibitem{SM18} 
   H. Sotani and U. Miyamoto, Phys. Rev. D {\bf 98}, 044017 (2018); {\bf 98}, 103019 (2018).

\bibitem{Sotani20} 
   H. Sotani, Phys. Rev. D {\bf 101}, 063013 (2020).

\bibitem{Riley19} 
   T. E. Riley {\it et al.}, Astrophys. J.  {\bf 887}, L21 (2019).
   
\bibitem{Miller19} 
   M. C. Miller {\it et al.}, Astrophys. J.  {\bf 887}, L24 (2019).
   

\bibitem{GNHL2011}
   M. Gearheart, W. G. Newton, J. Hooker, and B. -A. Li, Mon. Not. R. Astron. Soc. {\bf 418}, 2343 (2011).
   
\bibitem{SNIO2012}
   H. Sotani, K. Nakazato, K. Iida, and K. Oyamatsu, Phys. Rev. Lett. {\bf 108}, 201101 (2012);
   Mon. Not. R. Astron. Soc. {\bf 428}, L21 (2013); {\bf 434}, 2060 (2013).

\bibitem{SIO2016}
   H. Sotani, K. Iida, and K. Oyamatsu, New Astron. {\bf 43}, 80 (2016);
   Mon. Not. R. Astron. Soc. {\bf 464}, 3101 (2017); {\bf 479}, 4735 (2018); 
   {\bf 489}, 3022 (2019).
   
\bibitem{AK1996}
   N. Andersson and K. D. Kokkotas, Phys.\ Rev.\ Lett.\ {\bf 77}, 4134 (1996).

\bibitem{AK1998}
   N. Andersson and K. D. Kokkotas, Mon.\ Not.\ R. Astron.\ Soc.\ {\bf 299}, 1059 (1998).

\bibitem{STM2001}
   H. Sotani, K. Tominaga, and K. I. Maeda, Phys.\ Rev.\ D {\bf 65}, 024010 (2001).

\bibitem{SH2003}
   H. Sotani and T. Harada, Phys.\ Rev.\ D {\bf 68}, 024019 (2003);
   H. Sotani, K. Kohri, and T. Harada, {\it ibid}.\ {\bf 69}, 084008 (2004).

\bibitem{SYMT2011}
   H. Sotani, N. Yasutake, T. Maruyama, and T. Tatsumi, Phys.\ Rev.\ D {\bf 83} 024014 (2011).

\bibitem{PA2012}
   A. Passamonti and N. Andersson, Mon.\ Not.\ R. Astron.\ Soc.\ {\bf 419}, 638 (2012).

\bibitem{DGKK2013}
   D. D. Doneva, E. Gaertig, K. D. Kokkotas, and C. Kr\"{u}ger, Phys.\ Rev.\ D {\bf 88}, 044052 (2013).

\bibitem{FMP2003}
   V. Ferrari, G. Miniutti, and J. A. Pons, Mon. Not. R. Astron. Soc. {\bf 342}, 629 (2003).


\bibitem{FKAO2015}
   J. Fuller, H. Klion, E. Abdikamalov, and C. D. Ott, Mon.\ Not.\ R. Astron.\ Soc.\ {\bf 450}, 414 (2015).

\bibitem{ST2016}
   H. Sotani and T. Takiwaki, Phys.\ Rev.\ D {\bf 94}, 044043 (2016); {\bf 102}, 023028 (2020); arXiv:2008.00419.

\bibitem{SKTK2017}
   H. Sotani, T. Kuroda, T. Takiwaki, and K. Kotake, Phys.\ Rev.\ D {\bf 96}, 063005 (2017); {\bf 99}, 123024 (2019)

\bibitem{MRBV2018}
  V. Morozova, D. Radice, A. Burrows, and D. Vartanyan, Astrophys. J. {\bf 861}, 10 (2018).

\bibitem{TCPOF19}
   A. Torres-Forn\'{e}, P. Cerd\'{a}-Dur\'{a}n, A. Passamonti, M. Obergaulinger, and J. A. Font, Mon. Not. R. Astron. Soc. {\bf 482}, 3967 (2019).


\bibitem{SS2019}
   H. Sotani and K. Sumiyoshi, Phys.\ Rev.\ D {\bf 100}, 083008 (2019).


\bibitem{GW6}  
   B. P. Abbott {\it et al}. (LIGO Scientific and Virgo Collaborations), Phys. Rev. Lett. {\bf 119}, 161101 (2017).

\bibitem{EM}  
   B. P. Abbott {\it et al}. (LIGO Scientific and Virgo Collaborations), Astrophys. J. {\bf 848}, L12 (2017).


   

\bibitem{KS1999}
   K. D. Kokkotas and B. G. Schmidt, Living Rev. Relativ. {\bf 2}, 2 1999.
   
\bibitem{SIOO14} 
   H. Sotani, K. Iida, K. Oyamatsu, and A. Ohnishi, Prog. Theor. Exp. Phys. {\bf 2014}, 051E01 (2014).
   
\bibitem{SSB16} 
   H. O. Silva, H. Sotani, and E. Berti, Mon. Not. R. Astron. Soc. {\bf 459}, 4378 (2016).   

\bibitem{OI03} 
   K. Oyamatsu and K. Iida, Prog. Theor. Phys. {\bf 109}, 631 (2003).

\bibitem{OI07} 
   K. Oyamatsu and K. Iida, Phys. Rev. C {\bf 75}, 015801 (2007).

\bibitem{Shen} 
   H. Shen, H. Toki, K. Oyamatsu, and K. Sumiyoshi, Nucl. Phys. {\bf A637}, 435 (1998).


\bibitem{LP04} 
   J. M. Lattimer and M. Prakash, Science {\bf 304}, 536 (2004).
   
\bibitem{GCR12} 
   S. Gandolfi, J. Carlson, and S. Reddy, Phys. Rev. C {\bf 85}, 032801 (2012).

\bibitem{Annala18}  
   E. Annala, T. Gorda, A. Kurkela, and A. Vuorinen, Phys. Rev. Lett. {\bf 120}, 172703 (2018).

\bibitem{Sotani17}  
   H. Sotani, Phys. Rev. C {\bf 95}, 025802 (2017).

\bibitem{note}
The phenomena of avoided crossing has already been studied in terms of the radial oscillations of compact stars \cite{GL1999} and the nonradial oscillations of rotating neutron stars \cite{LS1996}. 

\bibitem{GL1999}  
   D. Gondek and J.L. Zdunik, Astron. Astrophys. {\bf 344}, 117 (1999).

\bibitem{LS1996}  
   U. Lee and T. E. Strohmayer, Astron. Astrophys. {\bf 311}, 155 (1996).

\bibitem{crust}
The ratio of the crust thickness, $\Delta R=R-R_c$, to the stellar radius is also characterized by the nuclear saturation parameters and stellar compactness \cite{SIO17}.

\bibitem{SIO17}
   H. Sotani, K. Iida, and K. Oyamatsu, Mon. Not. R. Astron. Soc. {\bf 470}, 4397 (2017).


\end{thebibliography}

\end{document}